\documentclass{aa}  

\usepackage{graphicx}
\usepackage[varg]{txfonts}
\usepackage{natbib}
\bibpunct{(}{)}{;}{a}{}{,}
\begin{document}

   \title{Molecules in the transition disk orbiting T Cha\thanks{Based 
on submillimeter and X-ray observations. Submillimeter observations have been collected at the European Organisation for Astronomical
Research in the Southern Hemisphere, Chile, with the Atacama Pathfinder Experiment APEX 
(Prog. ID 088.C-0441 and E-089.C-0518A). X-ray archival observations used in this paper have been obtained with {\it XMM-Newton}, an
ESA science mission with instruments and contributions directly funded by ESA member states and NASA.} \\}


   \author{G.G. Sacco
          \inst{1}
          \and
          J.H. Kastner\inst{2}
          \and
          T. Forveille\inst{3}
          \and 
           D. Principe\inst{2}
           \and
           R. Montez Jr.\inst{4}
           \and
           B. Zuckerman\inst{5}
           \and
           P. Hily-Blant\inst{3}
           }

   \institute{INAF-Osservatorio Astrofisico di Arcetri, Largo E. Fermi 5, 50125 Firenze, Italy\\
              \email{gsacco@arcetri.inaf.it}
         \and
             Center for Imaging Science and Laboratory for Multiwavelength Astrophysics, 
             Rochester Institute of Technology, 54 Lomb Memorial Drive, Rochester, NY 14623, USA
         \and    
             Laboratoire d'Astrophysique de Grenoble, Universit\'e Joseph Fourier-CNRS, BP 53, 
             38041 Grenoble Cedex France
         \and
             Department of Physics and Astronomy, Vanderbilt University, Nashville, TN 37235, USA
          \and
             Department of Physics and Astronomy, University of California Los Angeles 90095, CA, USA       
             }

   \date{Received ; accepted }

 
  \abstract
   {}
   {We seek to establish the presence and properties of gas in the circumstellar disk orbiting 
   T Cha, a nearby (d$\sim$110 pc), relatively evolved (age $\sim$5-7 Myr) yet actively accreting 
   1.5 M$_{\odot}$ T  Tauri star.}
   {We used the Atacama Pathfinder Experiment (\textit{APEX}) 12 m radiotelescope to search for 
submillimeter molecular emission from the T Cha disk, and we reanalyzed archival \textit{XMM-Newton} imaging 
spectroscopy of T Cha to ascertain the intervening absorption due to disk gas along the line of sight to the star ($N_H$).}
   {We detected submillimeter 
rotational transitions of $^{12}$CO, $^{13}$CO, HCN, CN and HCO$^{+}$ from the T Cha disk. 
The $^{12}$CO  line (and possibly the $^{13}$CO line) appears to display a double-peaked line profile 
indicative of Keplerian rotation; hence, these molecular line observations constitute the first direct 
demonstration of the presence of cold molecular gas orbiting T Cha. Analysis of the CO emission line data indicates that 
the disk around T Cha has a mass ($M_{disk,H_2} = 80~ \rm M_{\oplus}$) similar to, but more compact ($R_{disk, CO} 
\sim$ 80 AU) than, other nearby, evolved molecular disks (e.g. V4046 Sgr, TW Hya, MP Mus) in which cold molecular 
gas has been previously detected. The HCO$^{+}$/$^{13}$CO and HCN/$^{13}$CO,  
line ratios measured for T Cha appear similar to those of other evolved circumstellar disks (i.e. TW Hya and V4046 Sgr).
The CN/$^{13}$CO ratio appears somewhat weaker, but due to the low signal-to-noise ratio of our detection, this discrepancy 
is not strongly significant.   
Analysis of the \textit{XMM-Newton} X-ray spectroscopic data shows that the atomic absorption $N_H$ toward T Cha is 1-2 orders of magnitude 
larger than toward the other nearby T Tauri with evolved disks, which are seen with much lower inclination angles.
Furthermore, the ratio between atomic absorption and optical extinction $N_H/A_V$ toward T Cha is higher than 
the typical value observed for the interstellar medium and young stellar objects in the Orion Nebula Cluster.
This may suggest that the fraction of metals in the disk gas is higher than in the interstellar medium.
However, a X-ray absorption model appropriate for the physical and chemical 
conditions of a circumstellar disk is required to address this issue.}
   {Our results confirm that pre-main sequence stars older than $\sim$5 Myr, when accreting, retain cold molecular disks, 
   and that those relatively evolved disks display similar physical and chemical properties.}


   \maketitle
%

\section{Introduction}

Circumstellar disks serve both as sources of material for accreting 
stars and as the sites of nascent planetary systems. Observations that
can establish the physical conditions and evolution of the gaseous
components of such disks are essential to understand the accretion process and the
processes involved in planet formation. Observation of emission lines
from molecular species (e.g. CO, HCN, CN, HCO$^{+}$) in the
submillimeter represents a powerful tool for studying cold 
(10-100 K) gas located in the outer regions (R$>$10 AU) of
circumstellar disks. Submillimeter observations of
molecular emission from disks orbiting young stars have been carried
out in the last two decades, using both single dish and
interferometric facilities; many of these studies have focused on
relatively evolved pre-main sequence (pre-MS) star/disk systems that
are located away from dark clouds and, hence, are free of potential
contaminating cloud CO line emission (e.g. \citealt{Dutrey:1994,
  Dutrey:1997, Dutrey:2008, Kastner:1997, Kastner:2008, Thi:2004, Qi:2004,
  Qi:2006, Qi:2008, pietu:2007, Rodriguez:2010, Oberg:2010, Oberg:2011, Andrews:2012} and references therein). 

In the last few years, our group initiated a campaign of multiwavelength
observations of young, roughly solar-mass pre-main sequence (pre-MS)
stars within $\sim$100 pc that are still accreting gas from
their circumstellar disks.  Only four pre-MS stars (TW Hya, V4046 Sgr,
MP Mus and T Cha), all located in the southern hemisphere, are
known to share all of these properties. Thanks to their proximity and
ages, these stars are particularly suitable for studies of star and
planet formation processes: they are close enough for detailed study
of the spatial structure of their disks with high spatial resolution
facilities; they are old enough (ages $\sim$5-20 Myr) that their disks may already have spawned
giant protoplanets; and their disks still retain significant amounts of gas, as
demonstrated by signatures of stellar accretion in the optical through X-ray
regimes (e.g. \citealt{Alencar:2002,  Argiroffi:2012, Curran:2011}).

The archetype of these young stars is TW Hya. The presence of molecular gas orbiting TW Hya 
was first established via single-dish CO observations by \cite{Zuckerman:1995}; 
subsequently, its disk has been scrutinized via both 
single-dish molecular  line surveys \citep{Kastner:1997, Thi:2004} and 
interferometric imaging \citep{Qi:2004, Qi:2006, Qi:2008, Hughes:2011, Rosenfeld:2012}.
More recently, we detected molecular emission from disks orbiting two additional nearby, 
accreting pre-MS stars, V4046 Sgr and MP Mus \citep{Kastner:2008, Kastner:2010}. Like TW Hya, 
the former system has been investigated in multiple molecular tracers \citep{Kastner:2008, Oberg:2011} and has been mapped 
interferometrically \citep{Rodriguez:2010, Oberg:2011, Andrews:2012, Rosenfeld:2012a},
whereas thus far the MP Mus molecular disk has only been detected via single-dish spectroscopy of $^{12}$CO.
Given simple assumptions concerning disk CO abundance, the CO submillimeter emission and
mid-infrared and submillimeter continuum observations suggest gas-to-dust ratios close to unity for all three disks --- suggesting either that
these disks are have already depleted a large part of their primordial gas \citep[as initially proposed by][]{Dutrey:1997} or that the 
[CO]/H$_{2}$ number ratio is much smaller than the value of 10$^{-4}$ usually adopted for purposes of estimating molecular gas masses from CO data. 
Furthermore, certain radio molecular lines measured for V4046 Sgr and TW Hya suggest that 
the chemistry of the circumstellar gas is influenced by the strong high-energy (UV and/or X-ray) radiation fields of the stars \citep{Kastner:2008, Salter:2011}.

T Cha is a K0 V star of 1.5 M$_{\odot}$ that is likely a member of the $\epsilon$ Chamaeleontis Association, 
on the basis of its kinematic properties \citep{Torres:2008, Olofsson:2011, Murphy:2013}.
Kinematic data have been used to derive the distances to the Association and its members, 
with two recent studies finding distances in the range 107-110 pc, both to the Association and
T Cha itself \citep{Torres:2008, Murphy:2013}.
Hence, in this paper, we adopt the distance of 110 pc to T Cha. The age of the Association is more uncertain. 
\cite{Torres:2008} proposed an age of 6-7 Myr, while \cite{Murphy:2013} recently suggested a younger age (3-5 Myr). 
According to its position in the HR diagram relative to theoretical pre-MS sequence evolutionary tracks, 
T Cha appears to be older than the rest of the Association (10-12 Myr from \citealt{Kastner:2012} and \citealt{Murphy:2013}). However, \cite{Murphy:2013}
argue that evolutionary models imply ages for solar mass stars that are older than their actual ages due to a systematic error in the 
models or in inferred values of temperature and luminosity.

T Cha is characterized by highly variable optical brightness ($\sim$3 mag in V band) as well as broad
emission lines (e.g. H$\alpha$, H$\beta$, O I at 6300 \AA) indicative of active accretion onto the star \citep{Schisano:2009, Kastner:2012}. 
The variability of the optical magnitude, emission line intensities, and extinction measured toward the
star are likely associated with a circumstellar disk seen at relatively
high inclination angle; based on modeling of near-infrared interferometric data, \citet{Olofsson:2013} estimate that the disk inclination is $i\approx60^{\circ}$ 
(where $i=90^{\circ}$ would be edge-on). 
Such a disk viewing geometry is further supported by the relatively
large absorbing column of gas toward the
star that is inferred from X-ray spectroscopy ($N_H \approx 10^{22}$ cm$^{-2}$;
\citealt{Gudel:2010}). The spectral energy distribution (SED) of T Cha from optical to 
millimeter wavelengths has been studied by several authors \citep{Brown:2007, Olofsson:2011,
Olofsson:2013, Cieza:2011}. These studies indicate that the T Cha disk has an optically thick inner
disk (radius 0.13-0.17 AU) and an outer disk of radius $>10$ AU (see below) separated by a cavity. Infrared
adaptive optics imaging hints at the potential presence of a substellar (perhaps even planetary mass)
companion at $\sim$7 AU, which may be responsible for excavating the 
cavity in the T Cha dust disk \citep{Huelamo:2011}; however, a recent reanalysis of these
data indicates the excess infrared flux indicative of
this close companion may instead be
due to anisotropic scattering in the disk \citep{Olofsson:2013}.
An analysis of the SED from the far-IR to mm-wave shows that
the outer disk only extends from 10 to 30 AU, with very little mass outside
\citep{Cieza:2011}.  Line profiles of [Ne {\sc ii}] emission from T
Cha, obtained via high spectral resolution mid-IR spectroscopy,
indicate that the gaseous component of the inner disk is
photoevaporating due to high-energy irradiation by the star
\citep{Pascucci:2009, Sacco:2012}.

In this paper, we report the detection of submillimeter emission from the circumstellar disk orbiting T Cha 
in transitions of $^{12}$CO, its most abundant 
isotopologue $^{13}$CO, and three other trace molecular species (HCO$^{+}$, HCN, CN).
 In Sec.\ 2, we describe the new submillimeter observations; 
in Sec.\ 3 we describe the submillimeter data analysis as well as the
properties of 
X-ray emission from T Cha as
determined from a reanalysis of archival data; in Sec.\ 4 we discuss our results by comparing
the properties of the disk around T Cha with the other nearby transition disks TW Hya, V4046 Sgr and MP Mus; 
and in Sec.\ 5 we draw our conclusions.

\begin{table*}
\caption{Observation log}             
 \label{tab:obslog}      
\centering                          
\begin{tabular}{cccc}        
\hline\hline                 
 day             & Lines & Time & pwv \\
(dd-mm-yyyy)     &       & (h)        & (mm) \\
\hline                        
18-09-2011 & $^{12}$CO (3-2), $^{13}$CO (3-2)  & 4.2  & 0.6-0.8   \\
21-09-2011 & $^{12}$CO (3-2)                   & 2.2  & 0.3-0.9   \\
22-09-2011 & $^{12}$CO (3-2)                   & 2.2  & 0.3-0.9   \\
14-05-2012 & $^{13}$CO (3-2)                   & 2.7  & 0.8-1.1   \\
20-05-2012 & $^{13}$CO (3-2)                   & 5.2  & 0.6-1.1   \\
21-05-2012 & HCO$^{+}$ (4-3), HCN (4-3)        & 3.0  & 0.5-0.7   \\
27-07-2012 & HCO$^{+}$ (4-3), HCN (4-3)        & 5.3  & 0.7-1.0   \\
04-08-2012 & CN (3-2)                          & 2.2  & 1.4-1.8   \\
10-08-2012 & CN (3-2)                          & 2.8  & 1.0-1.2   \\
11-08-2012 & CN (3-2)                          & 2.0  & 0.9-1.8   \\
\hline                                   
\end{tabular}
\end{table*}


\section{Submillimeter Observations}

We observed T Cha (J2000 coordinates $\alpha$= 11:57:13.550, $\delta$= -79:21:31.54) 
with the Atacama Pathfinder Experiment (\textit{APEX}) 12 m telescope \citep{Gusten:2006} in service mode for a total time of 31.8 h (including overheads) over the course 
of 11 nights in 2011--2012 (Table \ref{tab:obslog}).
The first observation, in 2011 September, yielded detection of $^{12}$CO $J=3\rightarrow2$ emission; following this detection, in May, July and August 2012, 
we observed (and detected emission from) $^{13}$CO $J=3\rightarrow2$, HCO$^{+}~J=4\rightarrow3$, HCN $J=4\rightarrow3$ and CN $J=3\rightarrow2$.  

All observations used the SHFI/\textit{APEX}-2 receiver and XFFTS spectral backend. The half-power beamwidth and main-beam efficiency of the \textit{APEX} 12 m at 
the frequency range of the Table~\ref{tab:obslog} observations (330--357 GHz) are $\theta_{mb}\approx17''$ and  $\eta_{mb} \approx 0.73$, 
respectively\footnote{See http://www.apex-telescope.org/telescope/efficiency/.}.
During the first observing run in September 2011, we began by using beam-switching mode
with the wobbling secondary for background subtraction. These initial observations yielded detection of circumstellar 
CO from T Cha, but with an apparent strong, narrow $^{12}$CO ``absorption'' feature superimposed.

We determined that this narrow CO feature could be attributed to imperfect subtraction of 
emission from a background molecular cloud (Dcld 300.2–16.9; \citealt{Nehme:2008}). 
To properly subtract the emission of the cloud, we used position-switching mode during the second and third nights,  taking as reference positions 
four points offset $\sim$30$^{\arcsec}$ to the east, west, north and south of the position of T Cha. This approach allowed us to effectively
subtract the cloud emission from the spectrum of circumstellar CO emission. Furthermore, to 
measure the emission from the cloud, we obtained position-switched spectra using a distant
off-source reference point located well outside the compact cloud. 
The emission from the cloud is well fitted with a gaussian profile at central velocity $v_{cloud}=4.67 \pm 0.02~\rm km~s^{-1}$ (with respect to
the local standard of rest, LSR), with a peak temperature $T_{peak}=0.82\pm 0.07~ \rm K$ and $\sigma_{cloud}=0.23\pm 0.02~ \rm km~s^{-1}$.   
Emission from the cloud did not affect the observations of the other molecular transitions so, during the run performed in 2012, 
we used beam-switching mode for background subtraction.

\section{Data Analysis \label{sec:analysis}}

\subsection{Molecular line emission \label{sec:sub_analysis}}

To reduce and analyze the data, we used the CLASS\footnote{See http://iram.fr/IRAMFR/GILDAS/} 
radio spectral line data reduction package and our own IDL-based analysis tools. Specifically, we used CLASS to co-add spectral scans, 
correct for the beam efficiency,  and subtract baselines, while
line fitting was performed by IDL scripts. 

\begin{figure}
\includegraphics[angle=0, width=\hsize]{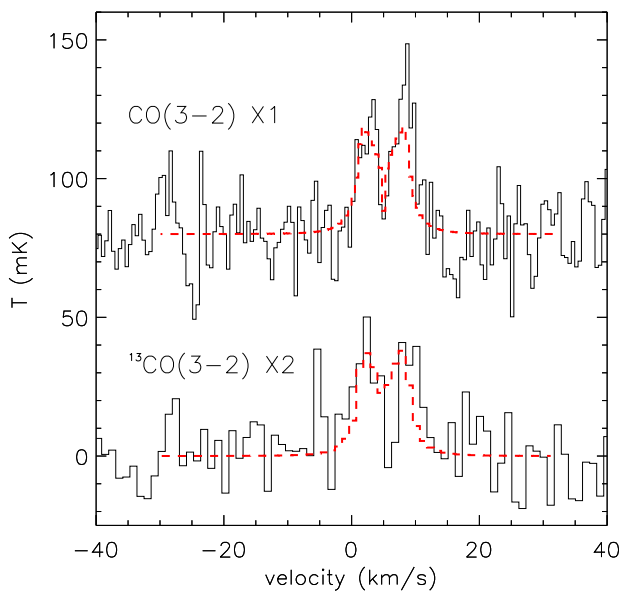}
\caption{Observed emission in the $^{12}$CO (3-2) and $^{13}$CO (3-2) transitions from T Cha.
The best fits of the lines with a Keplerian disk model profile are shown with red dashed curves.
The $^{13}$CO (3-2) intensity is multiplied by a factor 2 and the $^{12}$CO (3-2) baseline is offset in
temperature to allow a better comparison of the line profiles. \label{fig:CO}}
\end{figure}

\begin{figure}
\includegraphics[angle=0, width=\hsize]{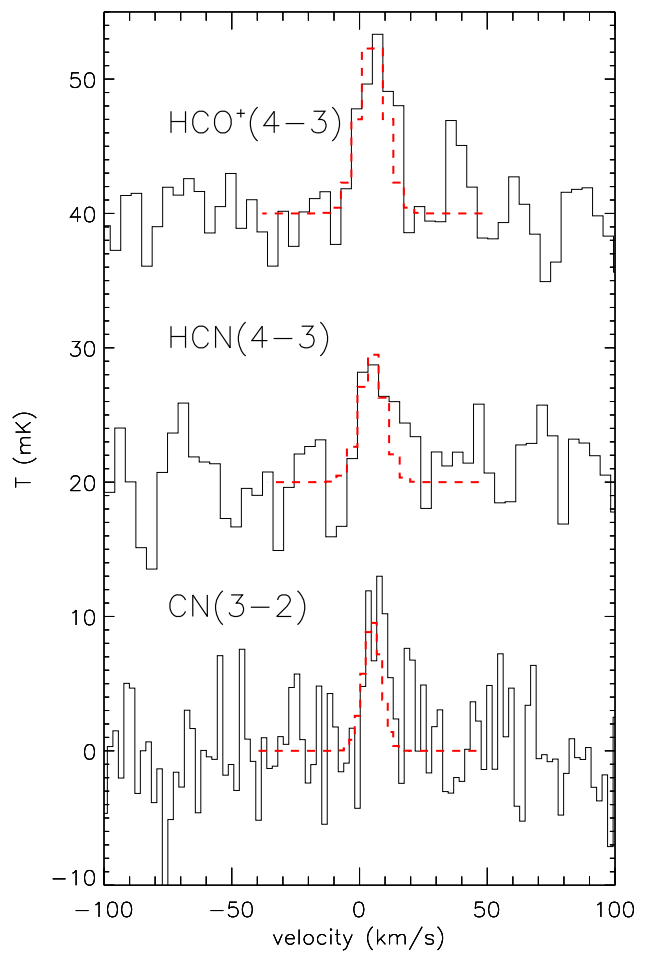}
\caption{Observed emission in the HCO$^{+}$ (4-3), HCN (4-3) and CN (3-2) transitions from T Cha.
The best fits of the lines with a gaussian profile are shown with red dashed curves.
The baseline of HCO$^{+}$ (4-3), HCN (4-3) are offset in temperature to allow a better comparison
of the line profiles. \label{fig:Other}}
\end{figure}

In Figures \ref{fig:CO} and \ref{fig:Other} we display the results for emission from the T Cha disk in the $^{12}$CO (3-2), $^{13}$CO (3-2), 
HCO$^{+}$ (4-3), HCN (4-3) and CN (3-2) transitions. 
Although somewhat noisy, the $^{12}$CO(3-2) line profile (and, possibly, the $^{13}$CO(3-2) line profile) appears to exhibit
steep sides and a double-peaked shape, as expected in the case of emission from an orbiting molecular disk (e.g. \citealt{Beckwith:1993}). Therefore, we fit the $^{12}$CO(3-2) line
with a parametric representation of the line profile predicted by a
Keplerian disk model, as
described in \cite{Beckwith:1993}. This parametric model was used by \cite{Kastner:2008, Kastner:2010} to analyze the
molecular emission detected from V4046 Sgr and MP Mus. The parameters of this model are: the peak temperature $T_{peak}$; 
the shift of the line centroid with respect to LSR, $v_0$;
the half-value of the velocity separation between the two line peaks $v_d$, which is equal to the 
radial velocity of the outer disk; the slope of the line wings $q$; and the slope of the 
central trough between the two peaks $p$. The $q$ parameter describes the disk radial temperature
profile $T\propto r^{-q}$, while $p$, in the case of an edge-on disk, indicates the definition 
of the disk outer edge (i.e., for a nearly edge-on disk, $p$=1 would correspond to a sharp outer edge, and a value $p<1$ indicates the lack of a sharp edge).

\begin{table*}
\caption{Results}
\label{tab:obstrans}
\centering
\begin{tabular}{ccccccc}
\hline\hline
Transition & $\nu$ & T$_{peak}$ & v$_{d}$           & $q$\tablefootmark{a} & p$_{d}$\tablefootmark{b} & Flux\tablefootmark{c} \\
           & (GHz) & (mK)       & ($\rm km~s^{-1}$) &     &         & ($\rm Jy~km~s^{-1}$) \\
\hline
$^{12}$CO (3-2) & 345.796000 & 38.1$\pm$2.3  & 3.5$\pm$0.2\tablefootmark{d} & 0.5 & 0.27$\pm$0.03          &   12.7$\pm$1.2  \\
$^{13}$CO (3-2) & 330.587960 & 20.0$\pm$5.4  & 3.5\tablefootmark{d,e}        & 0.5 & 0.27\tablefootmark{e}  &    7.0$\pm$2.1  \\
HCO$^{+}$ (4-3) & 356.734242 & 13.2$\pm$3.6  & 5.5$\pm$1.2\tablefootmark{f} &  -  &  -                     &    7.4$\pm$2.5  \\
HCN (4-3)       & 354.505469 &  9.5$\pm$3.9  & 4.9$\pm$1.6\tablefootmark{f} &  -  &  -                     &    4.9$\pm$2.5   \\
CN  (3-2)       & 340.247781 &  9.5$\pm$3.4  & 3.6$\pm$1.1\tablefootmark{f} &  -  &  -                     &    3.7$\pm$1.6   \\
\hline
\end{tabular}
\tablefoot{Results from the fit of the CO lines with a Keplerian disk model profile (see Sect. \ref{sec:analysis}) 
and of the other lines with a Gaussian.\\
\tablefoottext{a}{Power law index of the radial temperature profile within the disk. Fixed at a canonical value.}
\tablefoottext{b}{Parameter that describes the outer disk cutoff.}
\tablefoottext{c}{Converted from $\rm K~km~s^{-1}$ to $\rm Jy~km~s^{-1}$, using the conversion factor for \textit{APEX} reported
at the website http://www.apex-telescope.org/telescope/efficiency/}
\tablefoottext{d}{One half of the differences between red and blue peak velocities.}
\tablefoottext{e}{Assumed equal to the value derived from the fit of the $^{12}$CO (3-2) line.}
\tablefoottext{f}{$\sigma$ of the best fit Gaussian.}
}
\end{table*}

The best-fit parameters and the line fluxes determined from the model (with 1$\sigma$ errors) are reported in Table \ref{tab:obstrans}.
To fit the $^{12}$CO(3-2) profile, we left all parameters free with
the exception of the 
slope of the outer wings, $q$, which was fixed to the canonical value 0.5 \citep{Beckwith:1993};  to
fit the $^{13}$CO(3-2) line, all parameters with the exception of $T_{peak}$ were kept 
fixed at the values determined from the best fit to the $^{12}$CO(3-2)
profile. The velocity of the line centroid ($v_0=5.03\pm0.04$~km~s$^{-1}$, LSR) 
determined from the $^{12}$CO profile
is in agreement with the radial velocity of T Cha ($v_0=4.1\pm1.3~ \rm
km~s^{-1}$, after conversion to LSR), as
measured by \cite{Guenther:2007} via optical spectroscopy. However, our
measurement is more accurate, since multiple optical spectroscopic observations of
T Cha show that the radial velocity derived from the photospheric absorption lines
is variable due either the presence of a low-mass companion or very strong
stellar activity \citep{Schisano:2009}.
 
The value of $p$ is consistent with the values determined for the gaseous 
disks orbiting V4046 Sgr and MP Mus (\citealt{Kastner:2008, Kastner:2010}). 

Due to the weakness of the line fluxes with respect to the noise, it is not possible to determine reliable Keplerian model parameter 
values from the line profiles of the other three transitions observed (HCO$^{+}$, HCN and CN). 
Hence, we rebinned these data to coarser spectral resolution and fit 
all three lines with Gaussians whose central velocity was fixed to the  value determined from $^{12}$CO, i.e., 
$v_0 = 5.0~ \rm km~s^{-1}$. The resulting best-fit Gaussian parameters and line fluxes (with errors) for HCO$^+$, HCN, and CN are reported in Table \ref{tab:obstrans}.

\subsection{X-ray emission \label{sec:xray_analysis}}

\begin{figure}
\includegraphics[angle=0, width=\hsize]{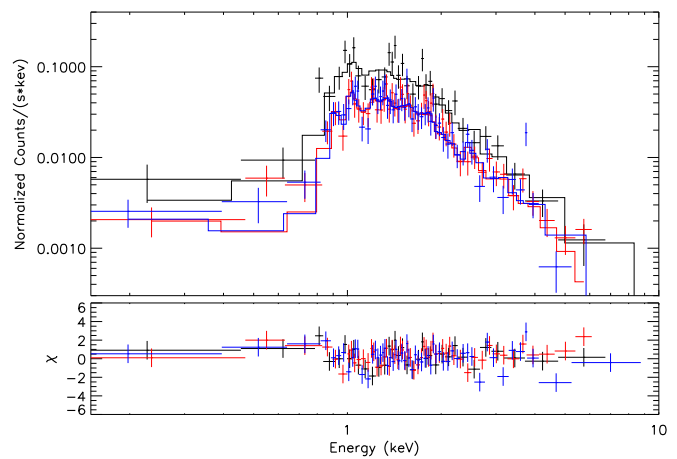}
\caption{The upper panel shows \textit{XMM-Newton} X-ray spectra (crosses) of T Cha filtered between 0.15-8.0 keV. The three colors 
(black, red and blue) represent the three detectors  
(pn, MOS1 and MOS2, respectively) of the instruments for low-resolution spectroscopy on board \textit{XMM-Newton}.
The histograms describe the best fit of the data with a model of the absorbed emission from a two temperatures optically thin plasma. 
The lower panel shows the residuals of the best fit.
\label{fig:X-ray}}
\end{figure}

T Cha was observed with \textit{XMM-Newton} on March 16, 2009 (ID 0550120601;
P.I. M. G\"udel). The star was well detected in this observation
\citep[as was a candidate wide-separation, low-mass
companion;][]{Kastner:2012}.  Cursory results of spectral analysis
performed on the XMM European Photon Imaging Camera (EPIC) detection
of T Cha --- specifically, an inferred intrinsic 0.3-10 keV X-ray luminosity of
$L_X = 1.1 \times 10^{30}$ erg s$^{-1}$ (assuming a distance of 66 pc) and intervening absorbing column of $N_H = 0.97  \times 10^{22}$
cm$^{-2}$ --- were reported by \cite{Gudel:2010}. Since T Cha lies
well in the foreground of the Cha dark clouds \citep{Torres:2008}, the
large value of $N_H$ determined by \cite{Gudel:2010} is evidently
dominated by gas in the disk along the line of sight to the star. In
light of our detection of cold molecular gas in the T Cha disk --- and
the likelihood that this cold gas is mainly responsible for the
absorption of soft X-rays from T Cha --- we have independently reduced
and analyzed the archival \textit{XMM-Newton} observation of T Cha, so as to
re-examine the determination of $N_H$ and its dependence on assumed
X-ray source spectral model parameters.

Standard SAS (v. 11.0) tasks were used to filter events and extract
EPIC pn, MOS1, and MOS2 spectra and spectral responses for T Cha.
The star was detected with all three instruments, with respective
count rates of 0.134, 0.064, and 0.062 counts s$^{-1}$ over effective exposure
times of 3.06, 10.07, and 10.67 ks. We used XSPEC (v. 12.6) to fit the
filtered 0.15--8.0 keV X-ray spectra with a model consisting of a
two-component, optically thin thermal plasma (XSPEC model
\verb+vapec+; \citealt{smith:2001, Foster:2012}) suffering intervening
absorption (model \verb+wabs+; \citealt{Morrison:1983}). Plasma
model abundances were initially uniformly set to
0.8 times solar; this uniform-abundance model yielded an acceptable
fit to the EPIC spectra ($\chi^{2} = 1.1$ for 140 degrees of
freedom) for parameters of $N_H = 1.2 \times 10^{22}~ \rm cm^{-2}$, $T_1=0.8$ $\rm keV$, 
$T_2=2.3$ $\rm keV$ and an intrinsic $L_{X} = 5.1 \times 10^{30}~ \rm erg~s^{-1}$ assuming a distance 
of 110 pc. However, this model fails
to reproduce a strong spectral feature near 1 keV
that would appear to be due to a blend of Ne {\sc ix} and Ne {\sc
  x} emission lines.

Hence, guided by the likelihood that the intrinsic X-ray spectrum of
the T Cha source bears a strong resemblance to those of the weakly
accreting (and more nearly pole-on) transition disk objects TW Hya
\citep{Kastner:2002,Brickhouse:2010} and V4046 Sgr
\citep{Gunther:2006}, we chose a model abundance pattern determined
for the former star's X-ray source \citep[specifically, Model C in
][]{Brickhouse:2010}. The pn and MOS spectra, overlaid with this
best-fit model, are displayed in Fig.~\ref{fig:X-ray}. We find
best-fit temperatures of $T_1=0.30_{-0.02}^{+0.04}$ keV and
$T_2=1.8_{-0.3}^{+0.4}$ keV and an intervening absorbing column of
$N_H = 1.97_{-0.15}^{+0.14} \times 10^{22}~\rm cm^{-2}$.  It would appear that the T Cha EPIC
spectra --- including the $\sim$1 keV feature --- are well described
by this heavily-absorbed but otherwise ``TW Hya-like'' model (reduced
$\chi^{2}=$ 1.1 for 140 degrees of freedom). As the precise values of
$T_1$ and $N_H$ are somewhat degenerate, their 90\% confidence ranges
were determined via a two-parameter $\chi^{2}$ analysis. Although this
analysis indicates that the best-fit values are relatively robust, it
is important to note that the fit results remain sensitive to the
assumed X-ray source atomic abundances (see also discussion in \S 4.3).

The foregoing best-fit two-component model with ``TW Hya-like''
abundances yields integrated (0.15--8.0 keV) observed (absorbed) and
intrinsic (unabsorbed) fluxes of $F_X=5.5_{-1.2}^{+0.4} \times
10^{-13}~\rm erg~s^{-1} cm^{-2}$ and $F_{X,0}=3.0 \times 10^{-11}~\rm
erg~s^{-1}~cm^{-2}$, respectively, where the large correction from
$F_X$ to $F_{X,0}$ is due to the combination of large $N_H$ and
relatively low soft-component temperature ($T_1$) required for the
best model fit. Adopting a distance of 110 pc for T Cha, the value
$F_{X,0}$ corresponds to an intrinsic X-ray luminosity of $L_{X} =
4.3 \times 10^{31} \rm ~erg~s^{-1}$, implying $\log{(L_{X}/L_{Bol})}=-2.43$ 
(given $\log{(L_{Bol}/L_{\odot})}=0.48$
from \citealt{Schisano:2009}, assuming a distance of 110 pc).

\section{Discussion \label{sec:discussion}}

\begin{table*}
\caption{Properties of Nearby T Tauri Star/Disk Systems}
\label{tab:nearbystars}
\centering
\begin{tabular}{lccccccccccc}
\hline\hline
Star & SpT & $M_\star$   & $D$  & age   &  incl.     & $R_{\rm disk, CO}$\tablefootmark{a} & $M_{\rm disk, CO}$\tablefootmark{b} &  $M_{\rm disk, H_2}$\tablefootmark{b} &$N_H$ & $M_{\rm disk, Dust}$&Ref \\
     &             & ($M_\odot$) & (pc) & (Myr) & ($^{\circ}$)    & (AU)  & ($M_\oplus$) & ($M_\oplus$) &  ($10^{20}cm^{-2}$) &($M_\oplus$) & \\
\hline
T Cha     & K0          & 1.5                   & 110     & 5-7   & 60    & 80   & 0.08                   & 80   & 197$\pm$15                                  &  7  &  1, 2, 3, 4 \\
MP Mus    & K1          & 1.2                   & 103     & 7     & 30    & 120  & 0.06\tablefootmark{c}  & 60   & 4.6$\pm$1.8\tablefootmark{d}                & 20  &  1, 5, 6, 7\\
V4046 Sgr & K5+K7       & 1.75\tablefootmark{e} & 73              & 12-20 & 35   & 370\tablefootmark{f}   & 0.1  & 100 & $\sim$2-4\tablefootmark{g}            & 20  &  8, 9, 10, 11, 12 \\
TW Hya    & K7          & 0.7                   & 54              & 8     & 6-7  & 200\tablefootmark{f}   & 0.02 & 20\tablefootmark{h} & 4-30\tablefootmark{i} & 60  &  13, 14, 15, 16 \\
\hline
\end{tabular}
\tablebib{(1) \cite{Torres:2008}; (2) \cite{Olofsson:2011}; (3) this work; (4) \cite{Cieza:2011}; (5) \cite{Kastner:2010} and refs.\ therein; 
(6) \cite{Argiroffi:2007}; (7) \cite{Carpenter:2005}; (8) \cite{Rodriguez:2010}; (9) \cite{Rosenfeld:2012a} and refs.\ therein; (10) \cite{Binks:2013}; 
(11) \cite{Kastner:2008}; (12) \cite{Argiroffi:2012}; (13) \cite{Rosenfeld:2012} and refs.\ therein;  (14) \cite{Thi:2004}; (15) \cite{Brickhouse:2010}; 
(16) \cite{Thi:2010}. }
\tablefoot{
\tablefoottext{a}{Derived from the fit of the $^{12}$CO line profile with a Keplerian disk model, using methods and assumptions described in Sect. \ref{sec:discussion_disk}, 
unless otherwise noted.}
\tablefoottext{b}{Calculated from the $^{13}$CO line flux, using methods and assumptions described in Sect. \ref{sec:discussion_disk}, 
unless otherwise noted.}
\tablefoottext{c}{Calculated from $^{12}$CO line flux, assuming the same flux ratio $^{12}$CO (3-2)/$^{13}$CO (3-2) observed for T Cha (i.e $\tau_{\rm 12CO} = 38$).}
\tablefoottext{d}{Derived from \textit{XMM-Newton} high resolution spectroscopic observations.}
\tablefoottext{e}{Total mass of central binary.}
\tablefoottext{f}{Disk CO radius based on interferometric mapping.}
\tablefoottext{g}{Range of variability within a one week monitoring.}
\tablefoottext{h}{\cite{Bergin:2013} estimated from the Hydrogen deuteride a disk mass 3 orders of magnitude larger than
our estimation from CO.}
\tablefoottext{i}{Range of values measured from different lines using high resolution \textit{Chandra} data.}
}
\end{table*}

\subsection{Disk structure, mass, and gas/dust ratio \label{sec:discussion_disk}}

Given some simple assumptions, high spectral resolution observations of circumstellar $^{12}$CO and
$^{13}$CO can be used to estimate the radius of the molecular disk,
and the mass contained in the disk
(see, e.g., \citealt{Zuckerman:2008, Kastner:2010} and references therein). 
Under the assumption of pure Keplerian rotation, we can derive the disk radius 
from the best-fit value for the outer disk radial velocity ($v_d$). 
Adopting a stellar mass $M=1.5~ \rm M_{\odot}$ \citep{Olofsson:2011} and an
inclination angle $i=60^{\circ}$ \citep{Olofsson:2013},
we estimate a disk radius $R_{\rm disk,CO}\sim80$ AU. As can be noted from 
Table \ref{tab:nearbystars}, this result for $R_{\rm disk,CO}$ is similar to (though 
somewhat smaller than) the radius estimated via similar CO line profile analysis for MP Mus
($\sim120$ AU), and is much smaller than the CO disk outer radii measured 
for V4046 Sgr and TW Hya ($\sim$370 and 200 AU, respectively) from interferometric
observations of CO \citep{Rodriguez:2010, Qi:2004, Rosenfeld:2012,
Rosenfeld:2012a}. However, the CO disk radius inferred for T Cha is 
not as small as its dust disk radius as determined from SED fitting, i.e., 
$R_{\rm disk,dust}\sim$30 AU (also assuming $i=60^\circ$; \citealt{Cieza:2011}).
This apparent discrepancy between molecular and dust disk
dimensions is similar to that found via interferometric imaging for other similarly evolved disks
(e.g., \citealt{Andrews:2012, Rodriguez:2010, Rosenfeld:2012a}). We note that
the radii of the molecular disks at V4046 Sgr and TW Hya estimated from the their CO line profiles
($\sim$250 and 165 AU from \citealt{Kastner:2008} and \citealt{Thi:2004}, respectively)
were both smaller than the radii subsequently measured from interferometers (see preceding paragraph), but still larger
than our estimate of the radius of the disk orbiting T Cha.

We estimated the disk mass from the $^{13}$CO line flux using equation
A9 in \cite{Scoville:1986}, which has been used to derive masses of
circumstellar disks by \cite{Zuckerman:2008} and \cite{Kastner:2008,
Kastner:2010}.  To obtain the disk CO mass, we assume 
negligible $^{13}$CO optical depth, a
CO excitation temperature $T_{ext}\sim 20~K$ \citep[as measured for the TW
Hya disk using interferometric observations of CO lines;][]{Qi:2004}, 
a distance $d=110~\rm pc$ \citep{Torres:2008}, and an
isotopic ratio $^{13}$CO/$^{12}$CO=69 \citep[appropriate for the local
ISM;][]{Wilson:1999}. To convert from CO to H$_2$ mass, we then adopt
a relative CO abundance of $\rm [CO]/H_{2}\sim
7\times 10^{-5}$ \citep[inferred for the Taurus molecular cloud;][ and references therein]{Dutrey:1997}.

The resulting disk H$_2$ gas mass, $M_{\rm disk,H_2}=2.4\times
10^{-4}~  M_{\sun}~(80~ M_{\oplus})$, corresponds to a gas to dust
mass ratio $\sim$12, adopting the dust mass
$M_{dust}=2.0\times10^{-5}~M_{\sun}$ derived by \cite{Cieza:2011} from
SED fitting. If compared with the canonical ISM value $\sim$100, this
result would indicate that the circumstellar gas has dissipated faster
than the dust component. However, we caution that the foregoing
(CO-based) estimated $ \rm H_2$ mass and gas-to-dust mass ratio may be
underestimated by a few orders of magnitude, given the uncertainties
associated with our many assumptions. Specifically: (a) $^{13}$CO may
not be optically thin; (b) the $^{13}$CO temperature may be lower than
20 K, as this temperature has been estimated from the optically thick
$^{12}$CO emission that traces the upper, warmer layers of the disk;
and (c) the CO abundance is likely $<7 \times 10^{-5}$, due to
photodissociation of CO or freeze-out of CO into dust grains.  

In Table \ref{tab:nearbystars}, we compare the foregoing results for
the disk gas mass of T Cha with the disk gas (CO and H$_2$) masses of
the other nearby (d$<$100) accreting T Tauri stars TW Hya, V4046
Sgr and MP Mus, as recalculated from single-dish measurements of their submillimeter CO
emission\footnote{For these disk mass calculations, 
we used the $^{13}$CO (3-2) and $^{13}$CO (2-1) line
fluxes for TW Hya and V4046 Sgr, respectively; while for MP Mus, we
estimated the $^{13}$CO (3-2) flux from the measured $^{12}$CO
(3-2) line flux, assuming the
same $^{13}$CO (3-2)/$^{12}$CO (3-2)$\sim 1.8$ flux ratio observed for
T Cha, which corresponds to an optical depth $\tau_{12CO}\sim 38$
under the assumption that the $^{13}$CO (3-2) emission is optically
thin.} using the same method and assumptions\footnote{Values
reported in Table \ref{tab:nearbystars} are slightly different from
the values reported in \cite{Kastner:1997, Kastner:2008,
Kastner:2010} due to different assumptions for gas
temperature, $^{12}$CO/$^{13}$CO isotopic ratio, CO optical thickness, and CO abundance.}
as for T Cha.  The disk around T Cha has a mass similar to those of the MP Mus
and V4046 Sgr disks and about 3.5 times the mass of the TW Hya disk
even though, as previously noted, the molecular disk radius estimated
from the T Cha line profile is smaller than in the case of the three
other, nearby T Tauri star disks listed in Table
\ref{tab:nearbystars}. Therefore, our observations of the gaseous
disk component appear to confirm that the disk orbiting T Cha is small
and dense, as suggested by \cite{Cieza:2011} based on their analysis
of mid- to far-IR continuum emission from dust. According to our
estimates, the TW Hya disk is less massive and, therefore, less dense
than the other disks; this is consistent with its $^{12}$CO optical depth,
$\tau_{\rm 12CO}\sim13$, which is smaller than the optical depths
estimated for T Cha and V4046 Sgr ($\tau_{\rm 12CO}\sim38$ and
$\tau_{\rm 12CO}\sim36$, respectively) using the same method. It is
interesting that the disk masses of the four stars are correlated
with the mass of the central stars, as already observed for the dust
masses on a much larger sample of stars \citep{Williams:2011}.
However, more accurate interferometric observations of $^{13}$CO
(3-2), $^{12}$CO and other isotopologues (i.e. C$^{18}$O and
C$^{17}$O) are required to better investigate the structures of these
circumstellar disks and thereby obtain more accurate estimates of their disk
gas masses. Indeed, recent observations of the disks orbiting TW Hya and V4046 Sgr,
combined with the results of detailed, self-consistent disk structure
and radiative transfer modeling, indicate that their disk H$_2$ gas masses (i.e.,
0.05--0.1 $M_\odot$; \citealt{Bergin:2013, Rosenfeld:2013}) are 2--3
orders of magnitude larger than the CO-line-based values.  
Furthermore, in both cases, there are indications of significant
variations in gas/dust mass ratio with disk radius
\citep{Andrews:2012, Rosenfeld:2013}.


\subsection{HCN, CN, HCO$^{+}$ vs. $^{13}$CO: comparison with other transition disks \label{sec:discussion_line_ratios}}

Our detection of T Cha in emission from HCN, CN, and HCO$^{+}$, in addition to
the two CO isotopologues, indicates that the chemical composition of
cold gas in the T Cha disk is similar to that of other, similarly
evolved disks. Specifically, the relative emission line fluxes we have
measured for T Cha in the 0.8 mm regime (see Table~2) are generally
similar to those measured for TW Hya and V4046 Sgr in the 1.3 mm and
0.8 mm regimes, respectively \citep{Kastner:1997,Kastner:2008}, with
the exception that T Cha appears to display somewhat weaker CN line
emission relative to $^{13}$CO (and, as noted by \citealt{Kastner:2008}, TW Hya displays
anomalously strong emission from HCO$^{+}$  relative to $^{13}$CO).

\cite{Kastner:2008} compared line ratios of HCN, CN, and HCO$^{+}$
emission with respect to $^{13}$CO for a small sample of (mostly)
isolated T Tauri disks, finding correlations among the three
ratios. These correlations were subsequently confirmed by
\cite{Salter:2011} on the basis of a larger sample of stars, including
young stellar objects in Taurus.  \cite{Kastner:2008} pointed out that
the relative abundances of HCN, CN, and HCO$^{+}$ are expected to be
enhanced in molecular gas that is irradiated by high-energy (ionizing)
photons; the fact that TW Hya and V4046 Sgr appear particularly strong
in all three of these tracers may then be indicative of their disks' cumulative
``doses'' of X-ray ionization, due to irradiation by the central stars
over their (relatively long) disk lifetimes. Although our results for T Cha
are less than definitive in this regard, due to the low
significance of our detections of $^{13}$CO, HCN, CN, and HCO$^{+}$,
it would appear that T Cha shows a similar pattern of enhanced HCN and
HCO$^{+}$ (if not CN) abundance, indicative of disk X-ray
irradiation. Clearly, additional, higher-quality measurements of
emission from the T Cha disk in these and other potential tracers of
disk irradiation are warranted.
 
\subsection{Implications of X-ray spectral analysis\label{sec:discussion_x-rays}}

In our X-ray spectral analysis (see Sect. \ref{sec:xray_analysis}), we confirm the basic
result, previously obtained by \cite{Gudel:2010}, that the T Cha X-ray source is
subject to an intervening absorbing column of order $N_H \sim 10^{22}$
cm$^{-2}$. We find, furthermore, that the inferred value of $N_H$ is not very
sensitive to the adopted intrinsic X-ray source model. 
This column density is much larger than the values of $N_H$ determined for
MP Mus, V4046 Sgr and TW Hya (see Table \ref{tab:nearbystars}).
This large discrepancy suggests that in stars harbouring disks seen at an high inclination angle, like T Cha 
($i \approx 60^\circ$), the molecular disk is the main contributor to the X-ray absorption,
while in the other stars which are viewed more nearly pole-on, like MP Mus, V4046 Sgr and TW Hya
(see Table \ref{tab:nearbystars}), atomic absorption can be 
due to material located much closer to the star (e.g. accretion columns connecting the inner
disk to the stellar photosphere).

The ratio between atomic absorption and optical extinction $(N_H/A_V)_{TCHA}$ 
lies in the range $\sim (4-16) \times 10^{21} \rm
cm^{-2}~mag^{-1}$ ($A_V \sim$1.2-4.6 mag). This is a factor $\sim$2--7 larger than
the ratio $(N_H/A_V)_{ISM}$ observed in the ISM ($(N_H/A_V)_{ISM}\sim 2.2\times 10^{21} \rm cm^{-2}~mag^{-1}$,
\citealt{Ryter:1996}), and larger than ratios observed for young 
stellar objects in the Orion Nebula Cluster \citep{Feigelson:2005}.  
The $(N_H/A_V)$ ratio depends on the dust grain properties, but
as discussed by \cite{Schisano:2009} and \cite{Cieza:2011}, the dust
grains within the T Cha disk are larger than ISM grains; hence, the extinction
curve is flatter than characteristic of the ISM (i.e., $R_{v}\sim 5.5$ for T
Cha; \citealt{Schisano:2009}) and, as a result, we
would expect the $(N_H/A_V)$ ratio to be lower than the ISM value
\citep{Draine:2003}. Thus, the relatively large value of $(N_H/A_V)_{TCHA}$ appears to
indicate that the fraction of metals in the gas phase is higher than in the dust phase, 
since metals (especially C, N, and O) are the main
contributors to X-ray absorption \citep{Morrison:1983, Vuong:2003}.

We caution, however, that the standard X-ray absorption model we and others employ to determine
$N_H$ has been developed for physical conditions appropriate to the
ISM \citep[e.g., 20\% of H in molecular form;][]{Wilms:2000}. Hence, application of this same
ISM-based model to the evolved circumstellar disk orbiting T Cha --- in
which the gas is likely predominantly molecular, and the gas/dust mass
ratio may vary significantly along the line of sight
\citep[e.g.,][]{Andrews:2012,Rosenfeld:2013} --- implies there may be large
systematic uncertainties in the results for $N_H$. Development of an X-ray
absorption model appropriate for the molecular-to-atomic gas fractions
and molecular abundances characteristic of circumstellar disks would
reduce these uncertainties, although such an effort is clearly beyond the scope of
this paper.

Significantly, given the assumption that the abundance patterns in
X-ray-emitting plasma are ``TW Hya-like'', the X-ray spectral model
fitting provides evidence for the presence of a soft plasma component,
with characteristic temperature $T_X \sim 3.5\times10^6$ K. A
similarly soft plasma component has been observed in several classical
(actively accreting) T Tauri stars --- most notably (for present
purposes), the other three stars in Table 3, i.e., TW Hya, V4046 Sgr,
and MP Mus \citep{Kastner:2002, Argiroffi:2007, Argiroffi:2012}. As in
these cases, the presence of such a component in the T Cha X-ray
spectrum could be indicative of soft X-ray emission produced by shocks
at the base of accretion columns.  

On the other hand, as a consequence of the large inferred value of
$N_H$, the intrinsic X-ray luminosity implied by the presence of such
a soft component would make T Cha unusually X-ray luminous among T
Tauri stars. Specifically, our model fitting implies
$\log{(L_{X}/L_{Bol})} = -2.43$, i.e., roughly an order of magnitude
larger than typical of T Tauri stars \citep[e.g.,][and refs.\
therein]{Kastner:2012}. Again, however, we caution that the inference
of luminous, soft X-ray emission from T Cha rests in large part on the
accuracy of the model describing X-ray absorption within its
circumstellar disk. Furthermore, as discussed in
\cite{Brickhouse:2010} and \cite{Sacco:2010}, any soft component
attributed to accretion shocks may be
affected by chromospheric absorption, depending on the location of the post-shock
zone.  X-ray gratings spectroscopy observations of T
Cha are therefore required to more conclusively demonstrate the presence of
accretion-shock-generated X-ray emission from the star.

Interestingly, our X-ray spectral fitting results are very similar to those obtained 
by \cite{Skinner:2013} for the transition disk system LkCa 15 --- which, like T Cha, 
is viewed at relatively high inclination
\citep[$i\approx50^\circ$;][]{pietu:2007}. 
In particular, both the T Cha and LkCa 15 X-ray spectral 
analyses reveal evidence for a "cool'' ($T_X\sim3\times10^6$ K) 
plasma component that dominates the total 
X-ray flux but is heavily absorbed, due (presumably) to intervening disk material.

\section{Conclusions \label{sec:conclusions}}

We have performed a series of submm observations of T Cha 
with the \textit{APEX} 12 m radiotelescope, and we reanalyzed the available
\textit{XMM-Newton} archival X-ray data for this star, 
with the aim of studying the physical and chemical properties of its
circumstellar disk. We obtained the following main results:

\begin{enumerate}

\item We detected molecular emission from the T Cha disk, providing the first evidence 
for the presence of cold gas out to large radii from the star (i.e., $>$10 AU). Specifically, 
we detected and measured the fluxes of the $^{12}$CO (3-2), $^{13}$CO (3-2), HCO$^{+}$ 
(4-3), HCN (4-3) and CN (3-2). The $^{12}$CO (3-2) line profile (and,
possibly, $^{13}$CO (3-2) line profile) is double-peaked, indicative
of Keplerian rotation. 

\item T Cha joins TW Hya, V4046 Sgr, and MP Mus as 
the fourth nearby ($D\lesssim100$ pc) classical T Tauri star of roughly solar mass
and age of at least $\sim$5 Myr, that is known to harbour a molecular disk. Its detection in radio molecular line emission
further strengthens the connection between 
the presence of a cold, gaseous disk and signatures of stellar accretion, even in stars in an advanced
stage of the pre-main sequence phase \citep{Kastner:2010}. 

\item From a parametric fit of a Keplerian disk model line profile to
  the measured
  $^{12}$CO (3-2) line profile, we find an outer disk projected
  rotational velocity of $v_d=3.5$~km~s$^{-1}$. Under the assumption of
  pure Keplerian rotation and a disk inclination angle $i=60^{\circ}$,
  this value of $v_d$ implies an outer disk radius $R_{\rm disk, CO} \sim 80$ AU.
  This CO disk radius is smaller than the radii measured
  interferometrically for other, similar transition disks, such as TW
  Hya and V4046 Sgr. However, the CO radius we infer for the T Cha
  disk is significantly larger than the dust disk radius previously
  deduced from its infrared SED. Submillimeter interferometric
  observations of T Cha that can provide direct measurements of the
  disk's geometrical properties are clearly warranted.

\item From the $^{13}$CO (3-2) line flux, we estimate a total disk gas
  mass $M_{disk, H_2}\sim 80~ \rm M_{\oplus}$ and a gas-to-dust mass
  ratio $\sim 12$, where the latter is based on a disk dust mass
  estimate from the literature \citep{Cieza:2011}.  These values are
  similar to those obtained for other disks of similar age. However,
  we caution that such (single-dish, CO-based) estimates
  may suffer from large uncertainties, most of which may lead to
  severe underestimates in disk gas mass.

\item From a reanalysis of archival \textit{XMM-Newton} X-ray
  observations, we find T Cha has an intrinsic X-ray luminosity
  $L_{X}=4.3\times10^{31}~ \rm erg~s^{-1}$, with an intervening
  atomic absorbing column of $N_H= 2.0\times10^{22}~ \rm
  cm^{-2}$. The X-ray spectral analysis yields
  evidence for a strong soft component, possibly indicative of
  accretion shocks.
  The relatively large value of $N_H$ is indicative of 
  absorption due to intervening gas that resides in the (highly inclined)
  T Cha disk. The resulting inferred ratio between atomic absorption
  and visual extinction for the T Cha disk lies in the range $N_H/A_V
  \approx$ 4--16$\times 10^{21}~ \rm cm^{-2}$.  This is somewhat
  larger than the $N_H/A_V$ ratios characteristic of the ISM and
  star-forming clouds, indicating that the disk gas is rich in metals. 
  However, a X-ray absorption model appropriate for the physical and 
  chemical conditions of a circumstellar disk is required to address this issue.
  
\item The intensities of HCO$^{+}$ and HCN emission relative to
  $^{13}$CO measured for T Cha are similar to the relative HCO$^{+}$
  and HCN line intensities of the (similarly evolved) disks orbiting TW
  Hya and V4046 Sgr. The relative intensity of CN line
  emission appears somewhat weaker in the case of T Cha, but due to the low 
  signal-to-noise ratio of our detection, this discrepancy is not strongly significant. 
  Additional, more sensitive measurements of the T Cha disk in these tracers may
  clarify whether the disk displays chemical signatures of the
  ionizing effects of X-ray irradiation --- as would be expected given
  the clear indications, from XMM-Newton X-ray spectroscopy, of a
  large X-ray absorbing column due to intervening disk material.

\end{enumerate}

\begin{acknowledgements}

We would like to thank the anonymous referee for useful and constructive comments and S. Murphy for the 
discussion about the distance and the age of T Cha.
This publication is based on data acquired with the Atacama Pathfinder Experiment (APEX). 
APEX is a collaboration between the Max-Planck-Institut fur Radioastronomie, the 
European Southern Observatory, and the Onsala Space Observatory. This research was supported 
in part by U.S. National Science Foundation grant AST-1108950 to RIT.

\end{acknowledgements}

\bibliographystyle{aa}
\bibliography{/Users/sacco/LETTERATURA/bibtex_all}

\end{document}